

\magnification=1200
\baselineskip=19pt
\overfullrule=0pt

\line{\hfill TIFR-TH-92-56}
\line{\hfill IISc-CTS-92-9}
\line{\hfill October, 1992}

\vskip .3in

\centerline{\bf Non-linear Field Theory of a Frustrated Heisenberg
Spin Chain}

\vskip .2in
\centerline{Sumathi Rao\footnote*{E-mail: sumathi@tifrvax.bitnet}
\footnote{$^ \dagger $}{Permanent address: Institute of Physics,
Sachivalaya Marg, Bhubaneswar 751005, India}}
\centerline{Tata Institute of Fundamental Research}
\centerline{Homi Bhabha Road, Bombay 400005}
\centerline{India}

\vskip .1in
\centerline{and}

\vskip .1in
\centerline{Diptiman Sen\footnote{**}{E-mail: diptiman@cts.iisc.ernet.in}}
\centerline{Centre for Theoretical Studies}
\centerline{Indian Institute of Science}
\centerline{Bangalore 560012}
\centerline{India}

\vskip .3in
\noindent

\def\nn{nearest-neighbour}
\def\nnn{next-nearest-neighbour}
\def\H{Heisenberg}
\def\bsi{{\bf S}_i}
\def\bsj{{\bf S}_j}
\def\ijnn{<i,j>\in n.n}

\def\vphi{{\vec \phi}}
\def\bl{{\bf l}}
\def\bS{{\bf S}}
\def\a{\alpha}
\def\b{\beta}
\def\g{\gamma}
\def\eabg{\epsilon_{\a\b\g}}
\def\dxy{\delta(x-y)}
\def\p{\partial}
\def\MG{Majumdar-Ghosh}
\def\Ha{Hamiltonian}
\def\L{Lagrangian}
\def\uR{{\underline R}}

{\bf Abstract}

We derive a continuum field theory for the \MG\ model in the
large-$S$ limit, where the field takes values in the manifold of
the $SO(3)$ group. No topological term is induced in the action
and the cases for integer spin and half-integer spin appear to
be indistinguishable. A one-loop $\beta -$function calculation
indicates that the theory flows towards a strong coupling
(disordered) phase at long distances. This is verified in the
large-$N$ limit, where all excitations are shown to be massive.

\vfill
\eject

\noindent
{\bf 1. Introduction}

The study of quantum spin models - quantum antiferromagnets in
particular - has attracted a great deal of interest [~1~-~4~]~,
ever since it was recognised that the ground state properties of
the Heisenberg antiferromagnet (AFM) and its generalisations in
two dimensions could be relevant to high $T_c$ superconductors
[~5~]. One widely used method of analysis which is applicable in
the long-wavelength limit has been the continuum field theory
approach, wherein an approximate field theory is derived for the
quantum AFM in the large-$S$ limit and the properties of the
ground state are obtained using standard field theory
techniques.

For one-dimensional quantum spin chains, the field theory
approach led to some interesting and unexpected results. By
mapping the Heisenberg AFM to a non-linear sigma model defined
on the manifold of $S^2~$, Haldane [~1,~2~] argued that an
integer spin system would have a ground state with massive
excitations and exponential disorder, but a half-integer spin
system would have massless excitations and algebraic order. The
distinction between integer spins and half-integer spins was
shown to be caused by the presence of a topological term in the
action. In two dimensions, however, for the Heisenberg AFM on
bipartite lattices, the current understanding [~3~,~4~] is that
no topological term is induced in the effective long-distance
action and the system always exhibits long-range N$\acute e$el
order (at zero temperature) with massless excitations.

However, for frustrated spin systems in  both one and two
dimensions, the ground states have not yet been conclusively
determined. It has been plausibly argued that the frustrated
$J~-~J^\prime ~$ model (with both \nn\ and \nnn\ couplings) on
bipartite two-dimensional lattices would have a flux-phase
ground state [~6~]. Another example of a two-dimensional
frustrated model - the Heisenberg AFM on a triangular lattice -
has been studied [~7~] in the field theory limit. It was shown
there that no topological term is induced and no distinction
between integer and half-integer spins is observed.

In one dimension, the \MG\ model [~8~], which has both
\nn\ and \nnn\ couplings with a particular ratio of
their strengths, is a specific example of a frustrated spin system.
This model has been solved for $S ~=~ 1/2 ~$ and has been shown
to have doubly degenerate valence bond ground states with
massive excitations [~8~]. However, frustrated spin models in
one dimension have not yet been studied from a field theoretic
point of view. In this paper, we shall address ourselves to a
study of the
\MG\ model in the large-$S$, long-wavelength limit. We shall derive
an appropriate field theory for the model and analyse it. In Sec. 2,
we briefly review the results from the spin wave, field theory
and large-$N$ approaches to the unfrustrated \H\ AFM. In Sec. 3,
we study the
\MG\ model in detail. We derive a spin wave theory about the
classical ground state using Villain's approach [~9~] and show
that the model has three spin wave modes, with two of them
having the same velocity and the third having a higher one. We
derive a continuum field theory for the model in the large-$S$,
long-wavelength limit and discuss its symmetries. No topological
term is induced in the action. Hence the long-wavelength theory
appears to be indifferent to the distinction between integer and
half-integer spins. We then perform a one-loop $\beta-$function
analysis of the field theory which indicates that the system
flows towards a disordered phase at long distances. Finally, we
generalise the field theory to an appropriate large-$N$ field
theory and show that a saddle-point approximation indicates that
all the excitations are massive. In Sec. 4, we summarise the
evidence offered in the earlier sections and present our
conclusions.

\vfill
\eject

\noindent
{\bf 2. The Heisenberg Antiferromagnet}

In this section, we shall briefly review [~1~,~2~] the study of
the simplest spin chain - the  \nn\ \H\ AFM - given by the \Ha\
$$
H = ~J \sum_{\ijnn} \bsi\cdot\bsj
\eqno(2.1)
$$
with $J>0$. The classical ground state (N$\acute e$el state) of
this model has neighbouring spins antiparallel. The standard
spin-wave analysis can be carried out about this ground state
using the Holstein-Primakoff transformation [~10~]
$$
\eqalign{
S_i^z &= S - a_i^{\dagger} a_i \cr
S_i^- &= a_i^{\dagger} ~(2 S - a_i^{\dagger} a_i )^{1/2}
\approx \sqrt {2S} ~a_i^{\dagger}\cr
S_i^+ &= (2 S - a_i^{\dagger} a_i )^{1/2} ~a_i
\approx \sqrt {2S} ~a_i }
\eqno(2.2)
$$
where the second approximate equality is true in the large-$S$
limit.  Two spin-wave modes are found at $k=0$ and $k=\pi$
satisfying the relativistic dispersion $E = ck$ with $c = 2JSa$.
However, it is well-known [~11~] that there exists no long-range
order in one dimension, due to infra-red divergences in the
theory. Hence, a small fluctuation analysis about an ordered
state is plagued with infinities and breaks down in the
disordered phase.

The next approach is the continuum field theory approach
initiated by Haldane [~1~], which led to the startling result
that integer spins have massive modes and exponential disorder,
whereas half-integer spins have massless modes and algebraic
order. Here, however, we shall follow Affleck's [~2~] method to
obtain the field theory. Continuum fields $\vphi$ and $\bl$ are
defined as
$$
\eqalign{
\bl_{2i+1/2} ~&= ~{(\bS_{2i} + \bS_{2i+1})\over 2a} ~\equiv ~\bl(x)  \cr
{\rm and} \quad
\vphi_{2i+1/2} ~&= ~{(\bS_{2i} - \bS_{2i+1})\over 2S} ~\equiv ~\vphi(x) \cr}
\eqno(2.3)
$$
where $a$ is the lattice spacing. These variables satisfy the
identities $$
\eqalign{
\bl\cdot\vphi &= 0 \cr
{\rm and} \quad
\vphi^2 &= 1 + {1 \over S} ~- ~{{a^2 \bl^2} \over S^2 }}
\eqno(2.4)
$$
so that in the large-$S$ limit, $\vphi^2 = 1$, - $i.e.$, $\vphi$
takes values on the manifold of the sphere $S^2$. The
commutation relations for the spins imply that the fields
satisfy the commutation relations given by $$
\eqalign{
[ ~l_{\a}, ~l_{\b} ~] &= i \eabg ~\l_{\g} ~\dxy  \cr
[ ~l_{\a}, ~\phi_{\b} ~] &= i \eabg ~\phi_{\g} ~\dxy \cr
[ ~\phi_{\a}, ~\phi_{\b} ~] &= 0   }
\eqno(2.5)
$$
which indicates that $\bl(x)$ can be identified as the angular
momentum of the field $\vphi$ constructed as $\bl = \vphi \times
{\vec \pi}$, (where ${\vec \pi}$ is the canonically conjugate
momentum to $\vphi$). The \Ha\ in Eq. (2.1) can be rewritten in
terms of these fields and expanded in a derivative expansion. In
the long-wavelength limit, we only need to keep terms upto two
space-time derivatives. In fact, for $\bl$, since it is already
proportional to ${\vec \pi}$, we only need to keep terms upto
$\bl^2$, ${\dot \bl}$ and $\bl'$, where the dot and prime denote
time and space derivatives respectively. In this limit, the
Hamiltonian can be reexpressed as
$$
H = \int dx ~[ ~{c g^2\over
2} ~( ~\bl + {S \over 2} ~\vphi^{~\prime} ~)^2 ~+~ {c\over 2
g^2} ~\vphi^{~ \prime ~2} ~]
\eqno(2.6)
$$
where $c = 2JSa$ is the spin-wave velocity and the coupling constant
$g^2 = 2/S$, so that large-$S$ corresponds to weak coupling.

This \Ha\ in Eq. (2.6) follows from the Lagrangian given by
$$
{\cal L} = ~{{\dot \vphi}^{~2} \over 2cg^2} ~-~ {c
\vphi^{~\prime ~2} \over 2g^2} ~+~ {S\over 2} \vphi \cdot
\vphi^{~\prime} \times {\dot \vphi}
\eqno(2.7)
$$
where the field $\vphi$ is subject to the constraint $\vphi^2 =
1$. Setting the spin-wave velocity $c=1$, the action can be
written in a relativistically invariant way as
$$
{\cal S} =
\int d^2 x ~[ ~{1\over 2g^2} ~\p_{\mu}\vphi \cdot \p^{\mu}\vphi
{}~+~ {\theta\over 8\pi}
\epsilon^{\mu\nu}\vphi \cdot \p_{\mu}\vphi \times \p_{\nu}\vphi ~]
\eqno(2.8)
$$
with $\theta = 2\pi S$. The $\theta$ term can actually be shown
to be a total derivative, which affects neither the classical
equations of motion nor the perturbative Feynman rules. However,
it does have a topological significance. On compactified
Euclidean space, both $x_{\mu}$ and $\vphi$ can be represented
as points on a sphere $S^2$.  The functional
$$
Q(\vphi) =
{1\over 8\pi}\int d^2 x ~\epsilon^{\mu\nu} 8~[\vphi \cdot
\p_{\mu}\vphi \times \p_{\nu}\vphi] ~\equiv ~Q
\eqno(2.9)
$$
measures the winding of the sphere $S^2(\vphi)$ onto the sphere
$S^2 (x_{\mu})$ and is always quantised as an integer for any
field configuration, since $\Pi_2 (S^2) = Z$. Thus, for integer
values of $S$, the contribution of the topological term to the
path integral is given by $e^{- 2\pi iSQ}=1$ - $i.e.$, it is
irrelevant. The action in Eq. (2.8) without the topological term
is well-known to describe a theory which flows to a strong
coupling, massive phase with masses of the order $e^{-\pi S}$.
But for half-integer spins, the topological term $e^{- 2\pi iSQ}
= e^{-i\pi Q}$ is either $+1$ or $-1$ depending on the value of
$Q$ and is certainly relevant.  In that case, the topological
term cannot be ignored. The current understanding [~2~] is that
the action in Eq. (2.8) with a topological term is actually
massless and describes a conformally invariant field theory with
algebraic order - $i.e.$, with a power-law fall-off of
correlation functions.

The result that the field theory in Eq. (2.8) describes a
massive phase can also be confirmed by generalising the $S^2$
manifold of the $\vphi$ field to an $S^N$ manifold and then
studying the large-$N$ limit of the field theory. The
appropriate \L\ is given by
$$
{\cal L} = ~{N\over 2g^2} ~[
{}~\p_{\mu}\vphi \cdot \p^{\mu}\vphi ~+~ i\lambda (\vphi^2 - 1) ~]
\eqno(2.10)
$$
where the constraint has been enforced using an explicit
Lagrange multiplier field. We can now integrate out $\vphi$ and
obtain
$$
{\cal S}_{\rm eff}(\lambda) = ~{N\over 2} ~{\Big[}
{}~-\int d^2 x ~{i\lambda
\over g^2} ~+~ {\rm tr}~{\rm ln}~(-\p^2 +i\lambda) ~{\Big]}~.
\eqno(2.11)
$$
In the large-$N$ limit, we can ignore fluctuations of $\lambda$
and evaluate ${\cal S}_{\rm eff}(\lambda)$ at the saddle-point
defined by $\p {\cal S}_{\rm eff}/\p\lambda = 0$. We find that
the saddle-point equation gives
$$
{1\over g^2} = {1\over 2\pi}
{\rm ln}~{\Lambda\over m}
\eqno(2.12)
$$
where $\Lambda$ is an ultra-violet cut-off. This determines the
mass parameter $m$ in terms of the cut-off and the bare
parameter $g$ as
$$
m = \Lambda e^{-2\pi/g^2} = \Lambda e^{-\pi S}
\eqno(2.13)
$$
which agrees with the expectation from the one-loop
renormalisation group flow.

Thus, the current understanding of the  \H\ AFM spin chain is
that both for integer and half-integer spins, there exists
short-range order, but for integer spins, at large distances,
long-wavelength fluctuations destroy the order. For half-integer
spins, even in the long-distance regime, excitations are
massless and there exists algebraic order.

\vfill
\eject

\noindent
{\bf 3. The \MG\ Model}

The Hamiltonian for the \MG\ model is given by
$$
H = ~J ~\sum_i ~\bsi \cdot ~[~ {\bf S}_{i + 1} ~+~ {1 \over 2}~
{\bf S}_{i + 2} ~]
\eqno(3.1)
$$
The classical ( $ S \rightarrow \infty~$) ground state must have
any three neighbouring spins adding upto zero. This can be seen
by recasting the \Ha\ in Eq. (3.1) as
$$
H = ~ {J \over 4} ~\sum_i ~( ~\bsi ~+~ {\bf S}_{i+1} ~+~ {\bf S}_{i+2} ~)^2
{}~+~ {\rm constant}
\eqno(3.2)
$$
For this to hold all along the chain, the spins must lie on a
plane and must successively twist by $~2 \pi /3 ~$ as shown in
Fig. 1.

For $S = 1 / 2 ~$, this model has dimerised ground states which
are just products of local singlet pairs. Since each spin can
form a singlet by pairing with its neighbour to the left or with
its neighbour to the right, it is clear that two such ground
states are possible. These states were proven to be exact ground
states by Majumdar and Ghosh. Later, it was shown [~12~] that
these two states are the only ground states and that there
exists a gap in the spectrum.

However, the model remains unsolved for all higher spins. Here
we shall obtain the ground state properties of the model in the
large-$S$, long-wavelength limit. Even though naive spin-wave
analysis is inadequate because of infra-red divergences in
one-dimensional systems, we perform a spin wave analysis to
obtain the spin wave modes and their velocities. The purpose of
this calculation is to have a check on the field theory
formulation. This is done in Sec. (A). In Sec. (B), we derive
the field theory, discuss its symmetries and show that it
reproduces the spin-wave calculations in the appropriate limit.
In Sec. (C), we study the renormalisation group flows of the
coupling constants. The global symmetries of the field theory
allow for four independent coupling constants. We study the flow
of these couplings by deriving the one-loop $\beta-$functions
and integrating them numerically. We obtain the scale at which a
transition to the strong coupling (disordered) phase occurs.
Finally, in Sec. (D), we show that in the large-$N$ limit, all
excitations become massive. This confirms the fact that the
theory is in the disordered phase in the long-wavelength limit.

\noindent
{\bf Sec. (A): Spin-wave Analysis}

Since the classical ground state of the \MG\ \Ha\ is planar, it
is more convenient to obtain the spin-wave theory by
parametrising [~13~] the spin operators in terms of Villain's
variables [~9~], rather than the usual Holstein-Primakoff
variables. The spin variables at each site are expressed in
terms of two conjugate variables $S_i^z $ and $\varphi_i ~$ as
$$
\eqalign{S_i^+ &= ~e^{i \varphi_i } ~{\Big[}~(~S + {1 \over 2} ~)^2 ~-~
(~  S_i^z + {1 \over 2} ~)^2 ~{\Big]}^{1/2} \cr
S_i^- &= ~{\Big[}~(~S + {1 \over 2} ~)^2 ~-~
(~ S_i^z + {1 \over 2} ~)^2 ~{\Big]}^{1/2}  ~e^{ - i \varphi_i }\cr
S_i^z &= {1 \over i} ~{\p \over {\p \varphi_i} } \cr}
\eqno(3.3)
$$
The periodic operator $\varphi_i ~(~\varphi_i ~=~\varphi_i ~+~ 2
\pi ~)$ satisfies the commutation relation
$$
[~\varphi_i ~,~
{S_j^z \over S} ~] ~=~ {i \over S} ~\delta_{ij} ~.
\eqno(3.4)
$$
In the large-$S$ limit, the discreteness of $S_i^z ~/ S$ and the
periodicity of $\varphi_i ~$ can be neglected and $S_i^z$ and
$\varphi_i$ can be thought of as canonically conjugate,
continuous momentum and position operators. In the classical (
$S \rightarrow \infty $ ) limit, in fact, $S_i^z ~/ S
\rightarrow 0$ and $\varphi_i ~$ is fixed at its classical value
${\bar \varphi_i}~$, which is the angle made by the classical
spin relative to the $x$-axis. The large-$S$ approximation
involves a systematic $1 / S $ expansion around this classical
ground state.

For the \MG\ \Ha, from Fig. 1, it is clear that ${\bar
\varphi_i} ~-~ {\bar \varphi_j} ~$ for any two neighbouring
spins is always $2 \pi / 3 ~$. Let us now compute the spin wave
spectrum for the \Ha\ by expanding the square roots in Eq. (3.3)
and keeping terms only upto order $1 / S ~$, and by expanding
the angles $\varphi_i ~$ to quadratic order about ${\bar
\varphi_i} ~$ - $i.e.$ we have
$$
\eqalign{
S_i^+ &\simeq ~e^{i {\bar \varphi_i}} ~(S ~+~ {1 \over 2}~)
{}~(1 + i \theta_i ~-~ {{\theta_i^2} \over 2} ~) ~{\Big(}1 ~-~ {1 \over 2}
{{(S_i^z + {1 \over 2} ~)^2 } \over {(S + {1 \over 2} ~)^2} } ~{\Big)}\cr
S_i^- &\simeq ~e^{-i {\bar \varphi_i}} ~(S ~+~ {1 \over 2}~)
{}~(1 - i \theta_i ~-~ {{\theta_i^2} \over 2} ~) ~{\Big(}1 ~-~ {1 \over 2}
{{(S_i^z + {1 \over 2} ~)^2 } \over {(S + {1 \over 2} ~)^2} } ~{\Big)}\cr}
\eqno(3.5)
$$
where $\theta_i ~$ describes in-plane fluctuations (fluctuations
about the angle ${\bar \varphi_i}~$) and $S_j^z / S ~$ describes
out-of-plane fluctuations. Substituting these operators in the
\Ha\ in Eq. (3.1), we find that
$$
\eqalign{
H = ~&J ~\sum_i ~[ ~ S_i^z ~ S_{i+1}^z ~-~  S_i^{z2} ~
{\rm cos} ~\Theta
{}~-~ {S^2 \over 2} ~(\theta_i ~- \theta_j ~)^2 ~ {\rm cos} ~\Theta ~] \cr
&+ ~{J\over 2} ~\sum_i ~[ ~ S_i^z ~ S_{i+2}^z ~-~ S_i^{z2} ~{\rm cos}
{}~2 \Theta ~-~ {S^2 \over 2} ~(\theta_i ~- \theta_j ~)^2 ~
{\rm cos} ~2 \Theta ~] \cr}
\eqno(3.6)
$$
where $\Theta = {\bar \varphi_i} - {\bar \varphi_j} = 2 \pi / 3 ~$.
Upon Fourier transforming, we can write the \Ha\ in the standard
oscillator form as
$$
H = J~[~\sum_q~ {S_q^z~ S_{-q}^z \over
2~m_q}~+~{1\over 2}~k_q~
\theta_q~\theta_{-q} ~]
\eqno(3.7)
$$
where
$$
\omega(~q~)~=~\sqrt{{k_q\over m_q}} ~=~-~ J~S~{\rm cos} ~\Theta~
\sqrt{~(3 - \g_q)~(3 - \g_q/{\rm cos} ~\Theta~)}
\eqno(3.8)
$$
with
$$
\g_q = 2~{\rm cos}~q ~+~ {\rm cos}~2q.
\eqno(3.9)
$$
Here, $-\pi<q< \pi~$. Also note that $\omega(q)~=~\omega(-q)~$.
The spin-wave spectrum is now obtained   by expanding
$\omega(q)$ about its zeroes -$i.e.$, about any $q_0$ where
$\omega(q_0) ~=~ 0~$ - and identifying its velocity as $\p
\omega / \p q~ \mid_{q=q_0}~$. It is clear from Eq. (3.8) that
$\omega(q)$ has three zeroes at $q~=~0, ~2\pi/3~$ and $-2\pi/3~$
within the first Brillouin zone. At $q~=~0~$, the spin-wave
velocity is given by $$ c~(q~=~0) ~=~ J~S~a~\sqrt{27\over 4}
\eqno(3.10)
$$
and at both $q~=~2\pi/3~$ and $q~=~-2\pi/3~$, it is given by
$$
c~(q~=~\pm~2\pi/3~) ~=~ J~S~a~\sqrt{{27\over 8}}.
\eqno(3.11)
$$
Hence, we have three spin-wave modes, two with the same velocity
and one with a higher velocity.

As was mentioned in Sec. 2, spin-wave theory about an ordered
state is not appropriate in one dimension, because infra-red
divergences in the theory drive any ordered state towards
disorder. However, the spin-wave velocities computed here will
serve as a consistency check on the field theory of the model,
which will be derived in the next subsection.

\vskip .2in

\noindent
{\bf Sec. (B): The Continuum Field Theory}

The first task in the derivation of a continuum field theory for
the \MG\ model is the identification of `small' and `large'
variables from a local set of spins, analogous to the $\bl$ and
$\vphi$ fields in Eq. (2.3).  Since, the classical ground state
is a three sub-lattice N$\acute e$el state, we define field
variables involving three local spins as
$$
\eqalign{
\bl_{3i}~&=~ {(~\bS_{3i-1} ~+~ \bS_{3i} ~+~ \bS_{3i+1}~)\over 3~a} \cr
(\vphi_{1})_{3i} ~&=~ {(~\bS_{3i-1} ~-~ \bS_{3i+1}~) \over \sqrt{3}~S} \cr
(\vphi_{2})_{3i} ~&=~ {(~\bS_{3i-1} ~+~
\bS_{3i+1} ~-~ 2~\bS_{3i}~)\over 3~S}}
\eqno(3.12)
$$
where $a$ is the lattice spacing. Note that we have grouped the
spins as $(3i-1,3i,3i+1)$. The two other possible groupings are
$(3i-2,3i-1,3i)$ and $(3i,3i+1,3i+2)$. All three groupings lead
to the same field theory.  (We shall verify this later and see
how these  three possibilities are related to symmetries in the
field theory.)

For the classical configuration in Fig. 1, we see that
$\bl~=~0~$ and the fields $\vphi_1$ and $\vphi_2$ satisfy the
conditions given by
$$
\vphi_1^2 ~=~ \vphi_2^2 ~=~1 \quad {\rm and} \quad \vphi_1\cdot\vphi_2 ~=~0.
\eqno(3.13)
$$
We may therefore expect $\bl~$ to be a `small' variable
(proportional to a space-time derivative in the long-distance
limit) and $\vphi_1~$ and $\vphi_2~$ to be the two `large' and
non-linear  variables, describing three degrees of freedom
(since $\vphi_1\cdot\vphi_2 ~=~0$). The actual identities
satisfied by these fields are given by
$$
\eqalign{
\vphi_1^2 ~+~\vphi_2^2 ~&=~ 2~+~{2\over S}~-~{2 a^2 \bl^2 \over S^2} \cr
\vphi_1^2 ~-~ \vphi_2^2 ~&=~ -~{4a\over S}~\vphi_2\cdot\bl \cr
\vphi_1\cdot\vphi_2 ~&=~ -~{2a\over S} \vphi_1\cdot\bl,   }
\eqno(3.14)
$$
which reduces to Eq. (3.13) in the long-distance (where $\mid
a\bl\mid~ <<~\mid\vphi_1\mid,~\mid\vphi_2\mid~$) and large-$S$
limit.  Moreover, the $\bl$, $\vphi_1$ and $\vphi_2$ fields
satisfy the commutation relations  given by
$$
\eqalign{
[~l_{\a}~(x), ~l_{\b}~(y)~] ~&=~ i~\eabg~l_{\g}(x)~\dxy \cr
[~l_{\a}~(x), \phi_{a\b}~(y)~] ~&=~ i~\eabg~\phi_{a\g}(x)~\dxy \cr
[~\phi_{a\a}~(x), \phi_{b\b}~(y)~] ~&=~ 0  }
\eqno(3.15)
$$
where $a,b$ take the values 1 and 2.
These relations can be easily derived from the commutation
relations of the spin operators by replacing the lattice
$\delta$-functions by continuum ones - $i.e.$, $\delta_{i,j}/3a
{}~\longrightarrow \dxy$.  Thus, $\bl$ can be identified as the
angular momentum operators of the fields $\vphi_1$ and
$\vphi_2$.

We can now derive the continuum \Ha\ by rewriting Eq. (3.1) in
terms of the fields $\bl$ , $\vphi_1$ and $\vphi_2$ and then
Taylor expanding the fields upto second derivatives of $\vphi_1$
and $\vphi_2$ and first derivatives of \bl, and by replacing the
sum over sites by an integral over space. Thus, in terms of the
fields, we obtain
$$ H = ~\int ~dx ~{\Big[}~ {{c g^2} \over 2}
{}~(~\bl ~+~ {{S {\vphi_1^{~ \prime}}}
\over {\sqrt 3}} ~)^2 ~+~ {c \over {2 g^2}} ~(~ {\vphi}_1^{~\prime ~2} ~+~
{\vphi}_2^{~\prime ~2} ~) ~{\Big]}
$$
where
$$
\eqalign{
c &= J S a ~(~{27 \over 8} ~)^{1/2} \cr
{\rm and} \quad g^2 &= {\sqrt 6} / S ~.}
\eqno(3.16)
$$
As before, we see that large-$S$ corresponds to the weak
coupling, perturbative regime. Since analysis of the field
theory is simpler from a
\L\ formulation, let us deduce a \L\ for this field theory.
We define a third unit vector
$$
{\vphi_3} = {\vphi_1} \times {\vphi_2}
$$
with
$$
{\vphi_1} \cdot {\vphi_3} = {\vphi_2} \cdot {\vphi_3} = 0 ~~~~,
{}~~~~{\vphi_3^2} = 1 ~.
\eqno(3.17)
$$
This field satisfies the commutation relations expressed in Eq.
(3.15) for $a, ~b ~=~ 3$ as well. Let us now  define an
orthogonal matrix $\uR$ whose entries are
$$
\uR = ~\left(\matrix{{\phi_{11}} &{\phi_{21}} &{\phi_{31}} \cr
{\phi_{12}} &{\phi_{22}} &{\phi_{23}} \cr
{\phi_{31}} &{\phi_{32}} &{\phi_{33}} \cr} \right) ~.
\eqno(3.18)
$$
Thus, the three columns of $\uR$ are given by the vectors
${\vphi_1}$, ${\vphi_2}$ and ${\vphi_3} ~$ respectively. Notice
that since $\uR$ is orthogonal $(~ \uR^T ~\uR = \uR ~\uR^T ~= I
{}~)$, it only has three independent degrees of freedom. If we now
introduce a diagonal matrix
$$
I_2 =~ \left(\matrix{1 &0 &0 \cr
0 &1 &0 \cr 0 &0 &0 \cr} \right) ~,
\eqno(3.19)
$$
it is clear that the second term in the \Ha\ in Eq. (3.16) is
proportional to ${\rm tr}~(\uR'^T\uR'I_2)~$. But the problem is
to figure out a time derivative term in the Lagrangian involving
the $\uR$ fields that would lead to the first term in the same
\Ha.

To figure this out, we note that our problem involves an
$SO(3)$-valued field at each space-time point. Hence, we need
the method of quantisation on a group manifold which is derived
in Ref. [~14~]. Following their work, we can show that the
required Lagrangian is given by
$$
{\cal L} = {1\over 4cg^2}
{}~{\rm tr} ~({\dot \uR}^T ~{\dot \uR}) ~-~ {c\over 2~g^2} {\rm
tr} ~(\uR'^T ~\uR'~I_2) ~+~ {S\over\sqrt {3}}
{}~\vphi_1\cdot\vphi'_1\times{\dot \vphi_1}.
\eqno(3.20)
$$
Let us verify that the \Ha\ in Eq. (3.16) can be derived from
this Lagrangian [~14~]. $\uR$ may be parametrised by three
numbers $\xi_{\a}$, which are local coordinates on the manifold
of $SO(3)$. Thus, the derivatives of $\uR$ are
$$
\p_{\mu} \uR ~=~ \p_{\mu} \xi_{\a} ~\p_{\a} \uR
\eqno(3.21)
$$
where $\p_{\a}~=~ \p/\p_{\a}~$. The momenta $\pi_{\a}$ canonical to
the $\xi_{\a}$ can be found from the Lagrangian as
$$
\pi_{\a} ~=~ {\p~{\cal L}\over \p~{\dot \xi}_{\a}} ~=~ {1\over 4cg^2}
{}~{\rm tr}~[~(\p_{\a} ~\uR^T)~{\dot \uR} ~+~ {\dot \uR}^T ~(\p_{\a}~\uR)~]
{}~+~ {S\over \sqrt{3}}~(\vphi_1\times \vphi'_1)~\cdot \p_{\a}\vphi_1
\eqno(3.22)
$$
leading to the \Ha\
$$
H ~=~ \int dx~[{1\over 4cg^2} ~{\rm tr} ~({\dot \uR}^T ~{\dot \uR})
   {} ~-~ {c\over 2g^2} {\rm tr} ~(\uR'^T ~\uR'~I_2)~].
\eqno(3.23)
$$
Now, the $\pi_{\a}~$ are locally defined momenta conjugate to
the locally defined coordinates $\xi_{\a}~$. The globally
defined momenta (conjugate to the globally defined coordinate
variables $\uR~$) are given by
$$
\l_{\a} ~=~ \pi_{\b}~N_{\b\a}
\eqno(3.24)
$$
where $~(\p_{\b}~\uR~)~N_{\b\a} ~=~ i~T_{\a}~\uR~$ and $T_{\a}~$
are the generators of $SO(3)~$, so that $(T_{\a})_{\b\g} ~=~
i~\eabg~$. Hence, we find that
$$
l_{\a} ~=~ {i\over 2cg^2}{\rm
tr} ~({\dot \uR}^T ~T_{\a} ~\uR) ~+~ {iS\over
\sqrt 3}~(\vphi_1\times \vphi_1^{~ \prime})\cdot (T_{\a}~\vphi_1).
\eqno(3.25)
$$
Here, we have used the fact that the first column of the matrix
$T_{\a}~\uR~$ is given by $T_{\a}~\vphi_1~$. The construction of
$l_{\a}$ guarantees that
$$
\eqalign{
[~l_{\a}~(x), ~l_{\b}~(y)~] ~&=~ i~\eabg~l_{\g}(x)~\dxy \cr
[~l_{\a}~(x), ~\uR~(y)~] ~&=~T_{\a}~\uR ~(x)~\dxy   }
\eqno(3.26)
$$
which is identical (in component notation) to Eq. (3.15).
Moreover, by substituting for ${\dot \uR}~$ in Eq. (3.23) in
terms of $l_{\a}~$ using Eq. (3.25), we can verify that the \Ha\
in Eq. (3.23) is identical to that in Eq. (3.16).

Let us now study the Lagrangian given in Eq. (3.20), which we
have just proved is the continuum field theory of the \MG\ model
in the large-$S$ limit. Notice that the last  term in Eq. (3.20)
is a total derivative, and therefore, has no effect
perturbatively. In fact, we shall now argue that this
topological term actually vanishes for all smooth field
configurations in Euclidean space, and has no non-perturbative
effect either.  To demonstrate this, we observe that the
Euclidean action (setting $c$ = 1) is given by
$$
S_E ~=~ \int~
d^2 x~\{ {1\over 4g^2} ~({\dot\vphi}_1^2 ~+~ {\dot\vphi}_2^2 ~+~
{\dot \vphi}_3^2~)~+~{1\over 2g^2}~( {\vphi}_1^{~\prime ~2} ~+~
{\vphi}_2^{~\prime ~2} ~) \} ~-~ {4\pi iS\over\sqrt
3}~Q~[\vphi_1] ,
\eqno(3.27)
$$
where $Q~[\vphi_1]$ is defined as $Q~[\vphi]~$ was defined in
Eq. (2.9).  Finiteness of $S_E$ implies that $\vphi_1$,
$\vphi_2$ and $\vphi_3$ must go to constants at infinity, so
that space-time is compactified to $S^2~$.  The functional
$Q~[\vphi_1]~$ must still be an integer since $\vphi_1$ also
lies on $S^2~$ ($\vphi^2 ~=~1~$). However, since $\vphi_1$ is
actually part of an $SO(3)$ matrix $\uR$ and $\Pi_2(SO(3))
{}~=~0~$, $Q[\vphi_1]$ must actually vanish. Let us demonstrate
this explicitly. We define a vector
$$
\vphi~(\theta) ~=~ \vphi_1~{\rm cos}~\theta ~+~ \vphi_2~{\rm sin}~\theta.
\eqno(3.28)
$$
Clearly, $Q~[\vphi(\theta)]~$ is also an integer, since $\vphi
(\theta) $ also takes values on $S^2 ~$. However, since $\vphi
(\pi) = ~-~\vphi (0) ~$, $Q[\vphi (\pi)] = ~-~Q[\vphi (0)] ~$.
Since $Q[\vphi]$ is restricted to integer values, it cannot
change when $\theta$ is continuously varied. Hence, the only
consistent possibility is that $Q[\vphi]$ must vanish for all
$\theta$.

Thus, we have the important result that there is no topological
term in the field theory of the \MG\ model in $1 ~+~ 1$
dimensions. Notice that the proof of the vanishing of the
topological term needed the global existence of two orthonormal
vectors ${\bf \phi}_1 ~$ and ${\bf \phi}_2 ~$, whereas in Sec.
2, the field theory of the \H\ AFM involved only one globally
defined field ${\bf \phi} ~$. Hence, the field theory defined in
Sec. 2 did possess a topological term which distinguished
between integer and half-integer   spins. Here, however, we can
work with the simpler $SO(3)$-valued field theoretic action
defined by
$$
S_E =~ \int ~d^2 x ~[~ {1 \over {4 g^2}} ~{\rm tr}
{}~({\dot \uR^T }  \uR) ~+~ {1 \over {2 g^2}} ~{\rm tr}
{}~(\uR^{~\prime T} ~\uR^\prime ~I_2)~] ~.
\eqno(3.29)
$$

This action is invariant under the global symmetry $\uR
\rightarrow A \uR B ~$, where $A$ is an $SO(3)$ matrix and $B$
is a block diagonal matrix of the form
$$
B = ~\left(\matrix{{\rm cos} ~\theta &{\rm sin} ~\theta &0 \cr -{\rm
sin} ~\theta &{\rm cos} ~\theta &0 \cr 0 &0 &1 \cr} \right)
\eqno(3.30)
$$
Thus, the theory has an $SO(3)_L \times SO(2)_R ~$ symmetry,
where the `L' index mixes the rows and the `R' index mixes the
columns.  The $SO(3)_L ~$ denotes the original spin symmetry of
the Hamiltonian generated by the angular momentum $\bl(x)$.
Equivalently, we may think  of this symmetry as the symmetry of
the `top' $\uR$ about a `space-fixed' set of axes that affects
the three vectors ${\vec \phi}_1 ~$, ${\vec \phi}_2 ~$ and
${\vec \phi}_3 ~$ in the same way without mixing them. On the
other hand, the $SO(2)_R ~$ is a rotation about the `body-fixed'
axis ${\vec \phi}_3 ~$ which mixes ${\vec \phi}_1 ~$ and ${\vec
\phi}_2 ~$. This symmetry is not present in the original spin
\Ha\ in Eq. (3.1) but only arises in the long-wavelength field
theory developed around the ground state of Fig.1. We now recall
the statement following Eq. (3.12).  To define the fields, the
spins can be grouped in threes in three different ways.  These
three ways are related to each other by rotations in the $(
{\vphi_1} , {\vphi_2} ~)$ plane by the angles $0, 2 \pi /3 ~$
and $ 4 \pi /3 ~$.
So we should certainly have expected a cyclic $Z_3 ~$ symmetry
in the field theory. However, the vanishing of the topological
term in the action appears to have allowed an enlargement of the
symmetry to a full continuous $SO(2)$ symmetry.

Finally, let us derive the spin-wave spectrum from the \L\ and
check whether we get the right spin-wave velocities. The
classical ground state is clearly given by the configuration
${\bS_{3 i}} ~= S ~(0, 1, 0), ~{\bS_{3 i - 1 }} ~= S ~(- {\sqrt
3} / 2, - 1 / 2, 0)~$ and ${\bS_{3 i + 1 }} ~= S ~({\sqrt 3} /
2, - 1 / 2, 0) $, which can be expressed in terms of the matrix
$\uR$ by the identity matrix i.e. $ \uR = I$.  Small
fluctuations about this ordered state can be parametrised by
$$
\uR = ~\left(\matrix{1 &- \g &\b \cr
\g &1 &- \a \cr
- \b &\a &1 \cr} \right)
\eqno(3.31)
$$
where  $\a~$, $\b$ and $\g$ are small fluctuation fields. In
terms of these fields, the \L\ is given by
$$
{\cal L} = ~{1
\over {2c g^2}} ~( {\dot \a^2} ~+~{\dot \b^2}~+~{\dot \g^2})~ -~
{c \over {2 g^2}} ~( \a^{\prime 2}  ~+~\b^{\prime 2} ~+~ 2
\g^{\prime 2} ~) ~.
\eqno(3.32)
$$
Thus, the $\a$ and $\b$ modes have the velocity $c$ and the
$\g$-mode has the velocity ${\sqrt 2} c $ in agreement with the
spin-wave velocities in Eq.  (3.10) and (3.11) since $c = J S a
{}~({27 / 8})^{1/2} ~$ from Eq. (3.16).  From the matrix $\uR$ in
Eq. (3.31), it is clear that the $\g$-mode describes
fluctuations  within the plane of the classical vectors
${\vphi_1} ~$ and ${\vphi_2} ~$ - $i.e.$, within the plane of
the classical spins.  The $\a$ and $\b$ modes describe
out-of-plane fluctuations.

Thus, we have proved that the field theory of the \MG\ \Ha\
reproduces the standard spin-wave results in the weak coupling
limit.

\vfill
\eject

\noindent
{\bf Sec. (C): One-loop {$\b$}-function and the flow of coupling constants}

In the previous section, we derived an effective low-energy
field theory for the \break
\MG\ model and showed that it reproduced the standard spin-wave
scenario by expanding about the classical, ordered ground state.
However, more generally, we can study the long-distance
properties of such a field theory by applying the
renormalisation group. The $SO(3)_L ~\times SO(2)_R ~$ global
symmetry of the action in Eq. (3.29) implies that at any scale,
the effective (Euclidean) \L\ can have four different coupling
constants and can be written as
$$
\eqalign{ {\cal L}_E = ~&(~{1
\over {2 g_1^2}} ~-~ {1 \over {4 g_2^2}} ~) {\rm tr}~({\dot
\uR^T}~ {\dot \uR}) ~+~(~{1 \over {2 g_2^2}} ~-~ {1 \over {2 g_1^2}} ~)
{\rm tr}~({\dot \uR^T}~ {\dot \uR} ~I_2 ) \cr
{}~&+~(~{1 \over {2 g_3^2}} ~-~ {1 \over {4
g_4^2}} ~) {\rm tr}~({\uR^{~\prime T}}~ {\uR^{~\prime}})
{}~+~(~{1 \over {2 g_4^2}} ~-~
{1 \over {2 g_3^2}} ~) {\rm tr}~({\uR^{~\prime T}}~ {\uR^{~\prime}} ~I_2)~.}
\eqno(3.33)
$$
At microscopic distances of the order of the lattice spacing
$a$, where we derived the field theory, we have
$$
g_1^2 ~=g_2^2 ~= g_3^2 ~= 2 g_4^2 ~= g^2 $$ with $$ g^2 ~= ~{\sqrt 6} / S ~.
\eqno(3.34)
$$
(Compare Eq. (3.29) and Eq. (3.33) ). But these values change as
we move to larger distance scales in accordance with the
renormalisation group equations. (Note that an $SO(3)_L \times
SO(3)_R ~$ symmetric \L\ would have had $g_1 ~= g_2 ~$ and $g_3
{}~= g_4 ~$). The unusual parametrisation in Eq. (3.33) has been
chosen for convenience in studying the evolution of the small
fluctuations $\a$, $\b$ and $\g$, so that the spin-wave
(Minkowski)
\L\ takes the simple form given by
$$
{\cal L} = ~{1 \over {2 g_1^2}} ~( {\dot \a^2} ~+~ {\dot \b^2} ~) ~-~
{1 \over {2 g_3^2}} ~( {\a^{\prime 2}} ~+~ {\b^{\prime 2}} ~) ~+~
{1 \over {2 g_2^2}} ~ {\dot \g^2} ~-~ {1 \over {2 g_4^2}} ~ {\g^{\prime 2}}~.
\eqno(3.35)
$$
 From this \L, we read off the velocity of the $\a$ and $\b$
modes to be $g_1 ~/ g_3 ~$ and that of the $\g$-mode to be
$g_2~/ g_4 ~$.

To study the flow of the coupling constants to the long distance
regime, we calculate the four $\b$-functions
$$
\b (g_i^2 ~) = ~{d \over dy} ~g_i^2 (y)
\eqno(3.36)
$$
where $y = {\rm ln} (L / a ~)$ is a measure of the distance scale $L$.
This is done using the background field formalism [~15~]. We
expand the field $\uR (x)$ as
$$
\uR (x) = ~\uR_0 (x) ~e^{i \eta_{\a} (x) T_{\a}}
\eqno(3.37)
$$
where $\uR_0 (x)$ is a slowly varying field, which we take to be
a solution of the Euler-Lagrange equations of motion of the \L\
in Eq. (3.33), and $\eta_{\a} (x)$ are rapidly varying fields.
We then integrate over the $\eta_{\a} ~$ fields to obtain an
effective action for $\uR_0 ~$, from which the $\b$-functions
are obtained. (The field $\uR (x)$ has to be expanded as $\uR
(x) = \uR_0 (x)~{\rm exp}( i \eta_{\a} (x)~T_{\a} ) ~$ rather
than $\uR (x) = ~{\rm exp}( i \eta_{\a} (x) ~T_{\a} ) ~\uR_0
(x)$ in order to maintain the $SO(3)_L ~\times SO(2)_R ~$
symmetry for the effective action for $\uR_0 ~$).

On expanding the right hand side of Eq. (3.33) to second order
in $\eta_{\a} (x)$ (the first order terms vanish because
$\uR_0~$ extremises the action), we find that the Euclidean
propagator $M_{\a \b} ({\bf k}) = \langle \eta_{\a} ({\bf k})
\eta_{\b} (- {\bf k}) \rangle $ in momentum space ${\bf k} = ( k_0 ~, k_1 ~)$
has a diagonal form with
$$
\eqalign{M_{11} = M_{22} &= {{g_1^2 ~g_3^2} \over {g_3^2 ~k_0^2 ~+~ g_1^2
{}~k_1^2 ~}} \cr {\rm and} \quad
M_{33} &= {{g_2^2 ~g_4^2} \over {g_4^2 ~k_0^2 ~+~ g_2^2 ~k_1^2 ~}} \cr}.
\eqno(3.38)
$$
In addition, there are two kinds of vertices between the
$\uR_0$-fields (denoted by dotted lines) and the $\eta$-fields
(denoted by solid lines) as shown in Figs. 2 (a) and 2 (b). The
vertex in Fig. 2 (a) has one derivative of $\uR_0 ~$ and one
derivative of $\eta$. (Since the momentum flowing along the
$\uR_0$ line is negligible compared to the momentum flowing
along the $\eta$ lines, we have shown the $\eta$-momenta to be
${\bf k}$ and $ - {\bf k}$ respectively). The Feynman rule for
this vertex is given by
$$
\eqalign{
\Gamma_{\a \b} = ~-i k_0 ~&{\Big[}~ ({1 \over g_1^2} ~-~
{1 \over {2 g_2^2}} ~)
{}~{\rm tr}~{\dot \uR_0^T} ~\uR_0 ~( ~T_{\a} ~T_{\b} ~-~ T_{\b} ~T_{\a} ~) \cr
&+~  ({1 \over g_2^2} ~-~ {1 \over { g_1^2}} ~)
{}~{\rm tr}~{\dot \uR_0^T} ~\uR_0 ~( ~T_{\a} ~I_2 ~T_{\b} ~-~ T_{\b} ~I_2
{}~T_{\a} ~)~ {\Big]} \cr
- i k_1 ~&{\Big[}~ ({1 \over g_3^2} ~-~ {1 \over {2 g_4^2}} ~)
{}~{\rm tr}~{\uR_0^{~\prime T}} ~\uR_0 ~( ~T_{\a} ~T_{\b} ~-~ T_{\b}
{}~T_{\a} ~) \cr
&+~ ({1 \over g_4^2} ~-~ {1 \over {g_3^2}} ~)
{}~{\rm tr}~{\uR_0^{~\prime T}} ~\uR_0 ~( ~T_{\a} ~I_2 ~T_{\b} ~-~ T_{\b} ~I_2
{}~T_{\a} ~)~{\Big]}~. }
\eqno(3.39)
$$
The vertex in Fig. 2(b) has two derivatives of $\uR_0 ~$ and no
derivatives of $\eta$, and its Feynman rule is given by
$$
\eqalign{ {\tilde \Gamma}_{\a\b} ~=~ &-~({1\over
g_2^2}~-~{1\over g_1^2}~) ~{\rm tr} ~\{{\dot \uR}_0^T ~{\dot
\uR}_0~ (T_{\a}~I_2~T_{\b} ~-~ {1\over 2} ~T_{\a}~T_{\b}~I_2 ~-~
{1\over 2}I_2~T_{\a}~T_{\b}~)\} \cr ~&-~ ({1\over
g_4^2}~-~{1\over g_3^2}~) ~{\rm tr} ~\{{\uR'}_0^T ~\uR'_0~
(T_{\a}~I_2~T_{\b} ~-~ {1\over 2} ~T_{\a}~T_{\b}~I_2 ~-~ {1\over
2}I_2~T_{\a}~T_{\b}~)\} ~. }
\eqno(3.40)
$$

The one-loop contribution to the effective action is therefore
obtained by evaluating the diagrams in Figs. 3 (a) and 3 (b),
which arise from the vertices in Figs. 2 (a) and 2 (b)
respectively.  Both the diagrams in Fig. 3 have symmetry factors
of $1/2~$. The momentum integrals are cut off at short distances
by the lattice spacing $a$ and at long distances by the length
scale $L$ at which we wish to evaluate the effective action -
$i.e.$, $L^{-1} ~\le~ k_0,k_1 ~\le~ a^{-1}~$. It is also
convenient to use identities like
$$
(~\vphi_1\cdot{\dot\vphi}_2~)^2 ~=~ {1\over 2}~(~{\dot
\vphi}_1^2~+~ {\dot \vphi}_2^2 ~-~ {\dot \vphi}_3^2~)
\eqno(3.41)
$$
which follow from the constraints in Eq. (3.13) and Eq. (3.17).
We then obtain the $\b$-functions given by
$$
\eqalign{
\b(g_1^2~) ~&=~ {g_1^4\over 2\pi}~[~{g_1^2~g_3~g_4\over g_2^2} ~
{2\over (g_1~g_4~+~g_2~g_3~)} ~+~g_1~g_3~({1\over g_1^2}~-~{1\over g_2^2}~)~]
\cr
\b(g_2^2~) ~&=~ {g_2^4\over 2\pi}~[~g_1^3~g_3~({2\over g_1^2} ~-~
{1\over g_2^2}~)^2 ~+~2~g_1~g_3~({1\over g_2^2}~-~{1\over g_1^2}~)~]
\cr
\b(g_3^2~) ~&=~ {g_3^4\over 2\pi}~[~{g_3^2~g_1~g_2\over g_4^2}~
{2\over (g_1~g_4~+~g_2~g_3~)} ~+~g_1~g_3~({1\over g_3^2}~-~{1\over g_4^2}~)~]
\cr
\b(g_4^2~) ~&=~ {g_4^4\over 2\pi}~[~g_3^3~g_1~({2\over g_3^2} ~-~
{1\over g_4^2}~)^2 ~+~2~g_1~g_3~({1\over g_4^2}~-~{1\over g_3^2}~)~]. }
\eqno(3.42)
$$

Let us now integrate the $\b$-functions numerically starting
from their values at $y~=~0$ given in Eq. (3.34).  The one-loop
$\b$-functions have the property that they are invariant under
the rescaling
$$
y ~\longrightarrow~\lambda~y \quad {\rm and}
\quad g_i^2 ~\longrightarrow g_i^2/\lambda
\eqno(3.43)
$$
which implies that the $\b$-functions for different values of
$g_i^2~$ (or equivalently $1/S~$) are related by a scaling law.
This also means that if one or more of the coupling constants go
to infinity at any length scale $y~=~y_0~$, then the product
$y_0~g^2~$ (where $g^2 = \sqrt{6}/S ~$), must be a number
independent of the value of $g^2$.

 From our numerical study of the renormalisation group flows, we
find that the coupling constants run in such a way that the
velocity of the $\g$-mode increases and the velocity of the $\a$
and $\b$ modes decreases.  (This behaviour was verified over a
large range of initial values starting from $g^2 ~=~ 0.001$
through $g^2 ~=~ 1.0~$).  Although all four couplings increase
monotonically, the coupling $g_2 ~$ goes to infinity first, and
this happens at a length scale given by
$$
y_0 ~g^2 ~= 4.5
\eqno(3.44)
$$
(At that point, the other three couplings have the values $g_1^2
{}~(y_0 ~) / g^2 ~= 3.6, ~g_3^2 ~(y_0 ~) / g^2 ~= 28$ and $g_4^2
{}~(y_0 ~) / g^2 ~= 12 ~$).  The blowing up of one or more of the
couplings is normally interpreted as a sign that the system
becomes disordered at that length scale - i.e. at
$$
L_0 ~\sim a
{}~e^{y_0} ~\sim a ~e^{ 4.5 / g^2} ~~\sim a ~e^{1.8 S }~.
\eqno(3.45)
$$
Beyond this length scale, the action  in Eq. (3.29) is no longer
valid and a new action (presumably describing massive
excitations and, perhaps, some massless excitations) must become
applicable. Although our calculation cannot be used to obtain
any  information other than the scale $L_0 ~$ at which the
perturbative calculations break down, it is interesting to
speculate on the possible scenarios at longer distances. We can
think of three possible scenarios.

\noindent
(a) All three modes $\a$, $\b$ and $\g$ become massive at $L_0 ~$, so that
there is a gap in the spectrum $\Delta \sim 1 / L_0 ~$.

\noindent
(b) Only the $\g$-mode becomes massive at $L_0 ~$. The $\a$ and $\b$ modes
remain massless until a longer distance $L_1 ~$ after which they too become
massive.

\noindent
(c) Only the $\a$ and $\b$ modes become massive at $L_0 ~$, but the $\g$-mode
becomes massive only at a longer distance $L_1 ~$.

\noindent
Clearly, much more analysis is needed to decide between the three
possibilities [~16~].

\vfill
\eject

\noindent
{\bf Sec. (D): A Large-N Approximation}

A non-perturbative method of studying the mass generation in any
field theory is to go to the large-$N~$ limit of an appropriate
generalisation of the model. For the $S^2$ model studied in Sec.
2, the generalisation to an $S^N$ model was obvious. Here,
however, the obvious generalisation of $\uR\in SO(3)~$ to
$\uR\in SO(N)~$ runs into problems. To study the $SO(N)$-valued
field theory , it is necessary to impose the constraints using
Lagrange multiplier fields. But this increases the number of
degrees of freedom from $N(N~-~1)/2~$ to $N^2~$, which does not
match asymptotically in the large-$N~$ limit.

To find a suitable generalisation, let us write the Euclidean
$SO(3)~$-valued field theory as
$$
{\cal L}_E ~=~ {1\over
2g^2}~[~{\dot\vphi}_1^{~2} ~+~{\dot\vphi}_2^{~2} ~-~
(~\vphi_1\cdot{\dot\vphi}_2~)^2 ~+~{\vphi}_1^{~\prime ~2}
{}~+~{\vphi}_2^{~
\prime ~2}~]
\eqno(3.46)
$$
where we have eliminated $~{\dot {\vphi_3^{~2}}} ~= {\dot {\vphi_1^{~2}}}
{}~+~ {\dot {\vphi_2^{~2}}} ~-~ 2 (~{\vphi_1} \cdot {\dot {\vphi_2}} ~)^2
{}~$, but we still have to impose the constraints ${\vphi_1^{~2}} ~=
{\vphi_2^{~2}} ~= 1$ and ${\vphi_1} \cdot {\vphi_2} ~= 0$. Note that
${\cal L}_E ~$ still exhibits the $SO(3) ~\times SO(2) ~$ symmetry. We can
now generalise the three dimensional vector fields ${\vphi_1}$ and
${\vphi_2}$  to $N$-dimensional vector fields and write the large-$N$ \L\
as
$$
{\cal L}_E ~= {N \over {2 g^2}} ~[~{\dot {\vphi_1^{~2}}} ~+~
{\dot {\vphi_2^{~2}}} ~-~ ({\vphi_1} \cdot {\dot {\vphi_2}} ~)^2 ~+~
{\vphi_1^{~\prime ~2}} ~+~ {\vphi_2^{~\prime ~2}} ~] ~.
\eqno(3.47)
$$
This \L\  has an $SO(N) \times SO(2)$ symmetry and since it still satisfies
the same three constraints, it has $2 N - 3 $ degrees of freedom. We introduce
three Lagrange multiplier fields to impose the constraints and a fourth field
to reduce the quartic term in (3.47) to a quadratic form. We then obtain
$$\eqalign{
{\cal L}_E ~= {N \over {2 g^2}} ~[~{\dot {\vphi_1^2}} ~&+~
{\dot {\vphi_2^2}} ~+~
{\vphi_1^{~\prime 2}} ~+~ {\vphi_2^{~\prime 2}} ~+~ i \lambda_1 ~(
{\vphi_1^2} ~-~ 1) ~+~ i \lambda_2 ~({\vphi_2^2} ~-~ 1) \cr
{}~&+~ 2 i \lambda_3
{}~{\vphi_1} \cdot {\vphi_2} ~+~ \lambda_4^2 ~+~ \lambda_4 ~({\vphi_1}
\cdot {\dot {\vphi_1}} ~-~ {\vphi_2} \cdot {\dot {\vphi_1}} ~) ~] ~.}
\eqno(3.48)
$$
(By integrating over the $\lambda_i ~$ fields in the path
integral, we get back the \L\ in Eq. (3.47) ). Now, by
integrating out the ${\vphi_i} ~$ fields, we compute the
effective action as a function of $\lambda_i ~$ as
$$
\eqalign{
e^{- S(\lambda_i ~)} ~&= ~\int {\cal D} {\vphi_1} {\cal D}
{\vphi_2} ~{\rm exp} ~[ - ~\int ~d^2 x ~{\cal L}_E ({\vphi_1}, {\vphi_2},
\lambda_i ~) ~] \cr
&= ~\int ~\prod_{i=1}^N {\cal D} {\phi_{1i}} {\cal D} {\phi_{2i}}
{}~{\rm exp} ~[ - ~\sum_{i=1}^N ~\int ~{{d^2 k}
\over {(2 \pi)^2}} ~M_i ~] }
\eqno(3.49)
$$
where
$$
M_i ~= ( {\phi_{1i}} ~(- k) ~ {\phi_{2i}} ~( k) ~)
{}~\left(\matrix{k^2 ~+~ i \lambda_1 &i \lambda_3 ~+~ i \lambda_4 ~k_0 \cr
i \lambda_4 ~k_0 &k^2 ~+~ i \lambda_2 \cr} \right)
\left(\matrix{&{\phi_{1i}} (k) \cr
&{\phi_{2i}} (- k) \cr } \right)
\eqno(3.50)
$$
so that
$$
\eqalign{
S(\lambda_i ~) &= ~{N \over 2} ~[ ~- ~i ~{{\lambda_1 } \over {g^2}} ~-~
{}~i ~{{\lambda_2 } \over {g^2}} ~]\cr ~
+ {N \over 2} ~ \int ~{{d^2 k} \over {(2 \pi)^2}} ~
 {\Big[} ~&{\rm ln} ~\{ ~k^2 ~+~ {i \over 2}
{}~(\lambda_1 + \lambda_2 ~) ~+~ i ~{\sqrt {~{ {(\lambda_1
- \lambda_2 ~)^2} \over 4} ~+~ (\lambda_3 + \lambda_4 ~k_0 ~)^2 ~} }
{}~\} \cr
+ ~&{\rm ln} ~\{ ~k^2 ~+~ {i \over 2} ~(\lambda_1 + \lambda_2 ~) ~-~ i ~
{\sqrt {~ {(\lambda_1
- \lambda_2 ~)^2 \over 4} ~+~ (\lambda_3 + \lambda_4 ~k_0 ~)^2 ~ }}
{}~\}  ~{\Big]} ~.}
\eqno(3.51)
$$
Because of the factor of $N$ in $S(\lambda_i ~)$, in the
large-$N$ limit, the integral over $\lambda_i ~$ in the path
integral is dominated by the saddle-point of $S(\lambda_i ~)$
given by ${\p S} / {\p \lambda_i} ~= 0$. This is found to be at
$\lambda_1 ~=\lambda_2 ~= - i ~m^2 ~$ and $\lambda_3 ~=
\lambda_4 ~= 0$ where the saddle-point equation gives
$$
{1 \over g^2} ~= ~\int ~{{d^2 k} \over {(2 \pi)^2}} ~{1 \over {(k^2 ~+~ m^2 ~)}
^2 } ~~~.
\eqno(3.52)
$$
Hence
$$
m = \Lambda ~{\rm exp} ~( - ~2 \pi / g^2 ~) = \Lambda ~{\rm exp}
{}~( - 2.57 ~S)
\eqno(3.53)
$$
where $\Lambda$ is the ultra-violet cut-off $a^{-1} ~$. We can
verify that this solution is actually a minimum by computing
$D_{ij} ~= {\p^2 S} / {\p \lambda_i ~\p \lambda_j } ~$ at the
saddle-point and showing that all its eigenvalues are positive.
In fact,
$$
D_{ij} ~=~ \left(\matrix{1/4\pi m^2 &0 &0 &0 \cr 0
&1/4\pi m^2 &0 &0  \cr 0 &0 &1/2\pi m^2 &0  \cr 0 &0 &0 &1/g^2
\cr} \right)
\eqno(3.54)
$$
at the saddle-point.  Substituting the values of $\lambda_i$ at
the saddle-point back in Eq. (3.48), we find that
$$
{\cal L}_E
{}~=~ {N\over 4g^2}~[~{\dot\vphi}_1^{~2} ~+~{\dot\vphi}_2^{~2}
{}~+~{\vphi}_1^{~\prime ~2} ~+~{\vphi}_2^{~\prime ~2} ~+~
m^2~(\vphi_1^{~2}~+~\vphi_2^{~2}~) ~-~ 2m^2~],
\eqno(3.55)
$$
so that all $2N$ fields have become massive.

Thus, the large-$N$ approximation confirms our earlier result
that the field theory of the large-$S$ limit of the \MG\ model
actually describes a disordered phase with a mass gap $m \sim
e^{-KS}$, where $K$ is some number of order one.

\vfill
\eject

\noindent
{\bf 4. Discussion and Outlook}

Our results in the previous section clearly indicate that the
\MG\ model is exponentially disordered beyond some distance scale
and has a mass gap for all large spins $S$. Thus, for all half-integer
spins, by applying the Lieb-Schultz-Mattis theorem [~17~], we see that
the ground state has to be doubly degenerate, with each state breaking
parity (defined as reflection about a site). For $S~=~1/2$, these ground
states are just the dimerised ground states mentioned at the beginning
of Sec. 3. For higher half-integer spins, the ground states are not easy
to visualise, although the theorem is known to hold. However, for integer
spins, the degeneracy or non-degeneracy of the ground state is still an
open question.

Can the procedure described in Secs. 2 and 3 be used to study
general spiral phases? The derivation of the field theory presented in
Secs. 2 and 3 crucially depended upon grouping together all the
spins in one period and then identifying `large' and `small' variables.
These became the non-linear fields and their canonical momenta in the
long-distance field theory. This procedure clearly fails for the
general spiral phase for which the classical ground state is
aperiodic.

\vskip .2in

\noindent
{\bf Acknowledgments}

We thank D. M. Gaitonde for comments and criticisms at all stages
of this work. One of us (S.R.) would also like to thank CTS, Bangalore
for hospitality during the course of this work.

\vfill
\eject

\noindent
{\bf References}

\item{\bf 1.}{F. D. M. Haldane, Phys. Rev. Lett. {\bf 50}, 1153 (1983);
     Phys. Lett.{\bf 93A}, 464 (1983).}

\item{\bf 2.} {I. Affleck, in {\it Fields, Strings and Critical Phenomena},
Les Houches, 1988, ed. E. Brezin and J. Zinn-Justin, (North Holland,
Amsterdam,1989); I. Affleck, Nucl. Phys. {\bf B265}, 409 (1986);
I. Affleck, J. Phys. Cond. Matt.{\bf 1}, 3047 (1989); R. Shankar and N. Read,
Nucl. Phys.  {\bf B 336}, 457 (1990).}

\item{\bf 3.} {T. Dombre and N. Read, Phys. Rev. {\bf B38}, 7181 (1988);
    E. Fradkin and M. Stone, Phys. Rev. {\bf B38}, 7215 (1988);
    X. G. Wen and A. Zee, Phys. Rev. Lett. {\bf 61}, 1025 (1988);
F. D. M. Haldane, Phys. Rev. Lett. {\bf 61}, 1029 (1988).}

\item{\bf 4.} { S. Chakravarty, B. I. Halperin and D. R. Nelson, Phys. Rev.
{\bf B39}, 2344 (1989).}

\item{\bf 5.} {J. G. Bednorz and K. A. M$\ddot u$ller, Z. Phys. {\bf B 64},
189 (1986).}

\item{\bf 6.} {I. Affleck and B. Marston, Phys. Rev. {\bf B 37}, 3774 (1988);
X. G. Wen, F. Wilczek and A. Zee, Phys. Rev. {\bf B 39}, 11413 (1989).}

\item{\bf 7.} {T. Dombre and N. Read., Phys. Rev. {\bf B39}, 6797 (1989).}

\item{\bf 8.} {C. K. Majumdar and D. K. Ghosh, J. Math. Phys. {\bf 10}, 1388
(1969); C. K. Majumdar, J. Phys. {\bf C3}, 911 (1970).}

\item{\bf 9.} {J. Villain, J. de Phys. (Paris) {\bf 35}, 27 (1974).}

\item{\bf 10.} {P. W. Anderson, Phys. Rev. {\bf 86}, 694 (1952) ;
M. J. Klein and R. S. Smith, ibid, {\bf 80}, 1111 (1950). }

\item{\bf 11.} {S. Coleman, Comm. Math. Phys. {\bf 31}, 259 (1973);
N. D. Mermin and H. Wagner, Phys. Rev. Lett. {\bf 17}, 1133 (1966).}

\item{\bf 12.}{I. Affleck, T. Kennedy, E. H. Lieb and H. Tasaki, Comm.
Math. Phys. {\bf 115}, 477 (1988).}

\item{\bf 13.}{H. Nishimori and S.J. Miyake, Prog. of Theor. Phys.
{\bf 73}, 18 (1985).}

\item{\bf 14.} {A. P. Balachandran, G. Marmo, B.-S. Skagerstam and A. Stern,
{\it Classical Topology and Quantum
States}, (World Scientific, Singapore, 1991); A. P. Balachandran, G. Marmo,
B.-S. Skagerstam and A. Stern,
{\it Gauge Theories and Fibre Bundles - Applications to Particle Dynamics},
Lecture Notes in Physics, {\bf 188} (Springer-Verlag, Berlin, 1983);
A. P. Balachandran, S. Borchardt and A. Stern, Phys. Rev. {\bf D17}, 3247
(1978). }

\item{\bf 15.}{ E. Br$\acute e$zin and J. Zinn-Justin, Phys. Rev. {\bf B14},
3110 (1976); A. M. Polyakov, Phys. Lett. {\bf B59}, 79 (1975).}

\item{\bf 16.}{S. Rao and D. Sen, work in progress.}

\item{\bf 17.}{E. H. Lieb, T. Schultz and D. J. Mattis, Ann. Phys. (N. Y.)
{\bf 16}, 407 (1961); I. Affleck and E. H. Lieb, Lett. Math. Phys.
{\bf 12}, 57 (1986).}

\vfill
\eject

\noindent
{\bf Figure Captions}

\item{\bf 1.}{Classical ground state of the \MG\ model.}

\item{\bf 2.}{Interaction vertices between $\uR_0 ~$ and $\eta$. The dotted
lines denote the $\uR_0 ~$ fields and the solid lines denote the $\eta$
fields.}

\item{\bf 3.}{One-loop effective action for $\uR_0 ~$  obtained by integrating
over the $\eta$ fields.}

\end